# Achieving higher photoabsorption than group III-V semiconductors in silicon using photon-trapping surface structures

Wayesh Qarony[1†], Ahmed S. Mayet[1†], Ekaterina Ponizovskaya Devine[2], Soroush Ghandiparsi[1], Cesar Bartolo-Perez[1], Ahasan Ahamed[1], Amita Rawat[1], Hasina H. Mamtaz[1], Toshishige Yamada[2,3], Shih-Yuan Wang[2], M. Saif Islam[1*]

[1]Electrical and Computer Engineering, University of California—Davis; Davis, CA 95618, USA.
[2]W&WSens Devices, Inc., 4546 El Camino; Los Altos, CA 94022, USA.
[3]Electrical and Computer Engineering, Baskin School of Engineering; University of California, Santa Cruz, CA 95064, USA.

[†] These authors contributed equally to this work

[*]Corresponding author. Email: sislam@ucdavis.edu

**Abstract**

The photosensitivity of silicon is inherently very low in the visible electromagnetic spectrum, and it drops rapidly beyond 800 nm in near-infrared wavelengths. Herein, we have experimentally demonstrated a technique utilizing photon-trapping surface structures to show a prodigious improvement of photoabsorption in one-micrometer-thin silicon, surpassing the inherent absorption efficiency of gallium arsenide for a broad spectrum. The photon-trapping structures allow the bending of normally incident light by almost ninety degrees to transform into laterally propagating modes along the silicon plane. Consequently, the propagation length of light increases, contributing to more than an order of magnitude improvement in absorption efficiency in photodetectors. This high absorption phenomenon is explained by FDTD analysis, where we show an enhanced photon density of states while substantially reducing the optical group velocity of light compared to silicon without photon-trapping structures, leading to significantly enhanced light-matter interactions. Our simulations also predict an enhanced absorption efficiency of photodetectors designed using 30 and 100-nanometer silicon thin films that are compatible with CMOS electronics. Despite a very thin absorption layer, such photon-trapping structures can enable high-efficiency and high-speed photodetectors needed in ultra-fast computer networks, data communication, and imaging systems with the potential to revolutionize on-chip logic and optoelectronic integration.

**Introduction**

Emerging applications in the cloud–computing, optical communication and interconnects [1–4], internet-of-thing (IoT) integrated communication networks [5,6], light detection and ranging (LiDAR) assisted applications, such as autonomous vehicles [7], security, surveillance, artificial intelligence [8,9], and low-light enabled imaging applications in life sciences [10–12] can be made economical, and versatile by designing complementary metal-oxide-semiconductor (CMOS) compatible, or silicon (Si) compatible photonic-systems. However, as an indirect bandgap material, silicon inherently exhibits a weaker absorption coefficient in comparison to group III-V compound semiconductors [13,14]. The absorption in Si shows a dramatic reduction in the near-infrared (NIR) wavelength spectrum—a wavelength spectrum range essential to plenty of aforementioned optoelectronic applications. Owing to its weak absorption, Si-based photonic devices demand a thick substrate layer to fully utilize the illuminated optical stimuli, e.g., for 95% photon absorption of 850 nm wavelength, a ~50 μm thick Si layer is required [15]. Despite an enhanced absorption in such a thick Si absorption layer, a long transit time of the generated electron-hole pair and its inherent low carrier mobilities result in only sub-par photodetectors that do not comply with the performance requirements of the emerging photonic systems. In contrast with Si, the most commonly used III-V compounds, such as GaAs and InP, exhibit 15× and 40× higher absorption, respectively, at a wavelength of 850 nm [13]. GaAs show ~90% absorption of 850 nm incident wavelength in merely a 2.5 μm thick layer. Attributing to its higher absorption coefficient and exceptionally high carrier mobilities [16], GaAs-based photodetectors exceed the performance expectations for the current and emerging optoelectronic systems. Such exceptional performance helps the GaAs, and their alloys thrive as the mainstream materials for numerous emerging photonic applications. However, a CMOS incompatibility in the fabrication process of such devices calls for a tedious and costly hybrid integration with CMOS electronics. Recently, despite its sub-par performance, foundries have introduced Si-based optoelectronic integrated circuits (OEIC) in the CMOS infrastructure [17,18], as a trade-off between performance and the manufacturing/integration cost. Therefore, innovative techniques to enhance light-matter interaction are crucial to designing photodetectors with diminishingly thin Si films used in modern CMOS processes.

Researchers are actively working on engineering photon-trapping solar cells and light-emitting diodes (LEDs) with diffusive surface textures to enable randomized light scattering and have demonstrated considerable enhancement in efficiency [15,19–23]. The most recent studies focus on theoretical simulations of the integrated photon-trapping structures in solar cells, offering distinctly high absorption efficiency by generating enhanced photonic density of states (DOS) and guided optical modes [24–30]. These simulations utilize the wave optics approach, which applies realistic Maxwell's physics as opposed to the geometric optics to simulate a thin-film solar cell and show an enhancement in the effective optical path length that absorbs close to or beyond the geometrical limit. Different wave optics photon-trapping schemes, such as photonic crystal structures [31,32], metallic nanoparticles [33], plasmonic surfaces [34,35], and grating [36], have been extensively investigated. Many of these theoretical investigations have revealed a very high optical absorption enhancement enabled by the photonic light-trapping structures in silicon solar cells at longer illumination wavelengths [26,37,38]. With the emergence of advanced optical device design and fabrication techniques, several experimental demonstrations of high optical absorption efficiency in solar cells [39,40] and ultrafast photodetectors have been realized [41,42–48]. It should be noted that the photodetectors usually operate under reverse bias conditions, while solar cells are designed to operate at maximum power points and used under zero bias. To circumvent the arduous device fabrication processes, researchers measured the quantum efficiencies using 1-T-R, where T and R are transmission and reflection of the illuminated light [20,21,24]. In contrast, the quantum efficiencies of photodetectors were directly calculated from the measured I-V characteristics.

Despite tremendous development in the fabrication facilities, such prodigious absorption enhancement in Si is only theoretically demonstrated and yet to be experimentally realized [21,49].

In this work, we report an experimental demonstration of the performance enhancement of Si-photodetector by incorporating photon-trapping micro and nano surface structures. We have fabricated metal-semiconductor-metal (MSM) photodetector on a 1 µm thin silicon layer and integrated periodic photon-trapping hole arrays. We have utilized CMOS-compatible processes to fabricate the photodetectors. To present a fair comparison, we have fabricated two sets of devices with and without a photon-trapping hole array. These hole arrays assist in diverting normally incident beams of light almost orthogonally and facilitate a lateral propagation of light. Such engineered surface profiling efficiently guides and effectively slows down the propagating light beam and results in a dramatic improvement in absorption efficiency. We demonstrate remarkable enhancement of 80%, 85%, and 65% in the absorption efficiency in the photon-trapping-equipped photodetectors for the NIR wavelength spectrum at 800, 850, and 905 nm, respectively. Further, we show a ~20% reduction in the device capacitance due to a reduced effective device volume of photon-trapping-equipped Si photodetectors that can further result in an ultrafast performance due to the reduction in RC time constant [50].

Furthermore, with the help of finite-difference time-domain (FDTD) simulations, we have shown that most of the propagating modes in Si with photon-trapping structures exhibit lower optical group velocity compared to that of the conventional Si layer without photon-trapping structures. This contributes to an enhanced light-matter interaction ensuring a higher absorption. This enhancement in the absorption is shown to be comparable to that of GaAs absorption. We have also shown that an equivalent performance enhancement can be achieved with 30 nm and 100 nm thin Si layers. The performance of such ultrathin Si-based photo-trapping-equipped photodetectors are intriguingly encouraging to fabricate ultrafast photodetectors in the existing CMOS foundry framework [51,52].

**Fabrication of the photodetectors on a 1 µm thin Si layer**

A thin silicon (Si) slab with a thickness of 1 µm integrated with a photon-trapping structure of cylindrical shape is designed to maximize light absorption. In this effort, the geometry is optimized for the lattice structure, diameter, period, and depth of the photon-trapping holes that are filled with air. The cross-sectional structure is schematically shown in Fig. 1(a), S2, and S6. To fabricate the photodetectors, we have exploited the standard CMOS-compatible infrastructure. We have opted for a mesa-based photosensor geometry where the photon trapping holes are patterned on the top mesa. The devices are fabricated on a silicon-on-insulator (SOI) substrate with a 1 µm thick active Si layer (the absorber layer). Fig. 1(b) showcases the optical microscopy images of the fabricated photodetectors for a range of mesa diameter, hole diameter ($d$), and periodicity ($p$) of holes. Detailed microscopic images of the fabricated device are shown in Fig. S3 of the supplementary information (SI). A firsthand confirmation of the wavelength selectivity of these photon trapping holes is evident from the different color spectrums revealed during microscopic imaging for different devices with different $d$, and $p$ (Fig. 1(b)). The scanning electron microscopy (SEM) images of the planar and photon-trapping photodetectors are shown in Fig. 1(c) and 1(d), respectively. The planar (without holes) photodetectors are used as a control device to benchmark the performance enhancement resulting from the hole-array introduction. The inset of Fig. 1(d) indicates that the patterned holes are circular. Interdigitated aluminum fingers with a thickness of 100 nm and width of 300 nm are sputtered on Si. Cylindrical photon-trapping holes with a $p$, $d$, and hole depth of 1300, 1000, and 600 nm, respectively, are etched in the Si active layer. Fabricated photodetectors are isolated from each other, and coplanar waveguides (CPW) were delineated for high-speed operation. The details of the fabrication method are discussed in the SI (Fig. S1) [44].

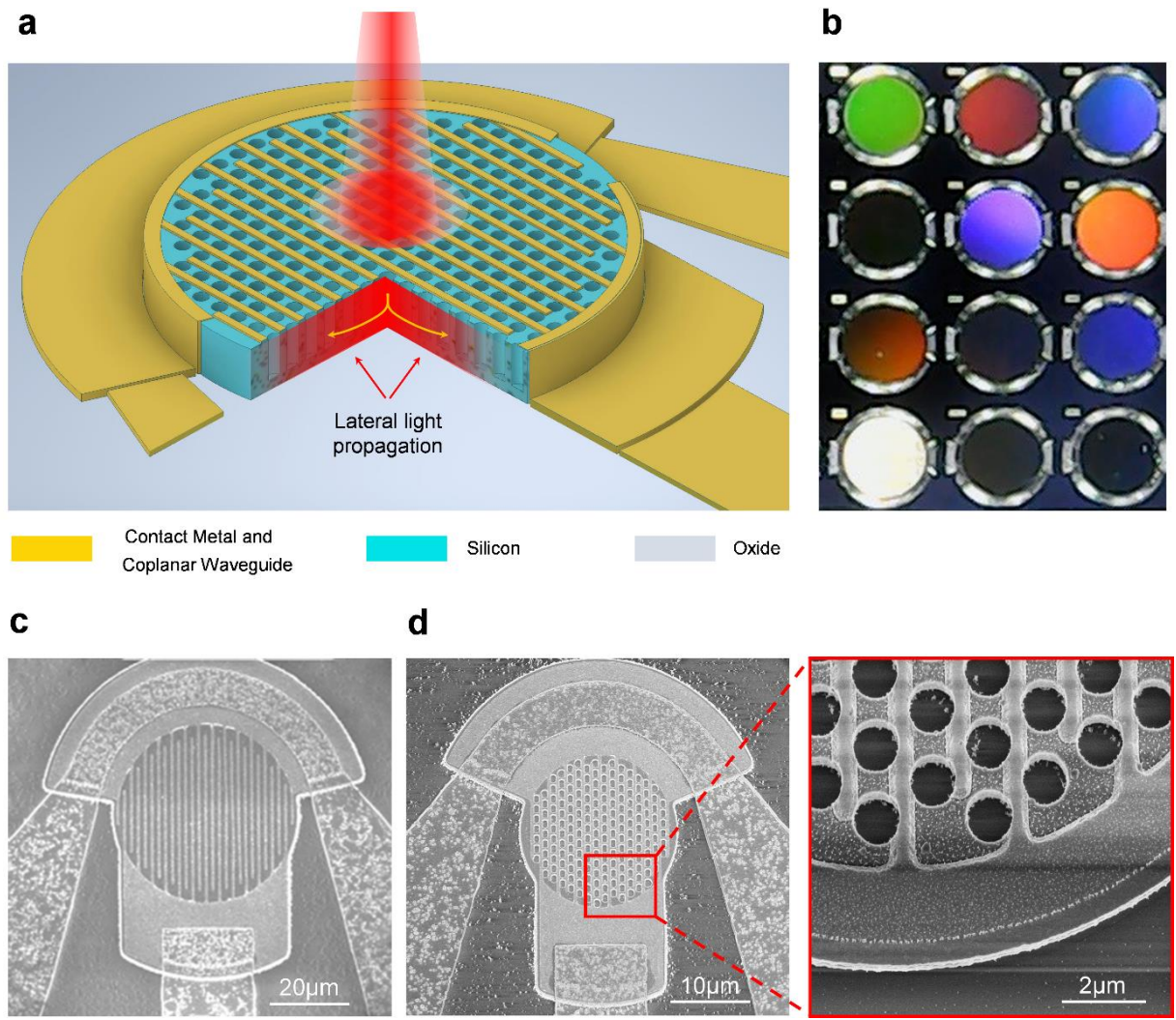

**Fig. 1. Design and fabrication of highly absorbing thin-film silicon photon-trapping photodetector.** (**a**) A Schematic of the photon-trapping silicon MSM photodetector. The photon-trapping cylindrical hole arrays allow lateral propagation by bending the incident light, resulting in an enhanced photon absorption in Si. (**b**) Optical microscopy images of the photon-trapping photodetectors fabricated on a 1 μm thin silicon layer of SOI substrate for a range of mesa diameters, *d*, and *p* of the holes. Under white light illuminations, the flat devices look white (bottom left) because of surface reflection. The most effective photon-trapping device looks black (bottom right). Less effective photon-trapping devices show different colors reflected from the surface of the devices. SEM images of fabricated (**c**) planar and (**d**) photon-trapping MSM photodetectors. The inset indicates circular shape holes in a hexagonal lattice formation.

**Leapfrogging the absorption coefficient of thin Si beyond that of intrinsic GaAs**

Using the Bouguer-Beer-Lambert law and considering surface reflection losses [53,54], an effective absorption coefficient ($\alpha_{eff}$), defined in Equation (1), is estimated to quantify the enhancement in photon absorption of the fabricated devices with photon-trapping structures.

$$\alpha_{eff}(\lambda) = -\frac{1}{t_{Si}}[ln(\frac{1-QE_{meas}(\lambda)}{1-R_{meas}(\lambda)})] \qquad (1)$$

Where $t_{Si}$ is the thickness of the Si active layer of the photosensor, and $QE_{meas}$, and $R_{meas}$ are the experimentally measured quantum efficiency and surface reflection of the devices respectively. If a perfect photon absorption is assumed with a quantum efficiency close to unity or 100% in a thin semiconductor film, the enhanced absorption efficiency ($\alpha_{eff}$) approaches an exceptionally high value. The measured surface reflections for the planar devices without anti-reflection coating range from 15–25%, whereas the surface reflections in the case of photon-trapping photodetectors show a notable reduction and are measured to be 10 – 12% for illumination wavelengths ranging from 800 to 905 nm. Using the experimental results and Equation (1), the $\alpha_{eff}$ of the Si photon-trapping photodetectors is estimated as a function of incident wavelengths. The estimated $\alpha_{eff}$ of the photon-trapping thin Si photodetectors is compared against the absorption coefficient of the bulk Si, and other potential photosensitive semiconductors such as Ge [55], InGaAs [55], GaAs [56], as shown in Fig. 2(a). The enhancement factor of the thin 1 μm thin photon-trapping Si photodetectors is not only higher than that of the bulk Si but also exceeds the absorption coefficient of GaAs over a broad near-infrared (NIR) wavelength spectrum and becomes comparable to the absorption coefficients of Ge and InGaAs. Quantitatively, in the photon-trapping hole-array equipped photodetectors with the measured quantum efficiencies of 80, 84, 86, and 68% at 800, 840, 850, and 905 nm respectively, $\alpha_{eff}$ is determined to be approximately 27467, 31797, 39713, and 14938 cm$^{-1}$. The $\alpha_{eff}$ determined at 850 nm is more than 70× and about 4× higher than the intrinsic absorption coefficient of Si (535 cm$^{-1}$) and GaAs (10,035 cm$^{-1}$) [56], respectively. Hence, the effective absorption coefficient of the fabricated photon-trapping photodetectors exceeded the intrinsic absorption coefficient of GaAs in the NIR wavelength region.

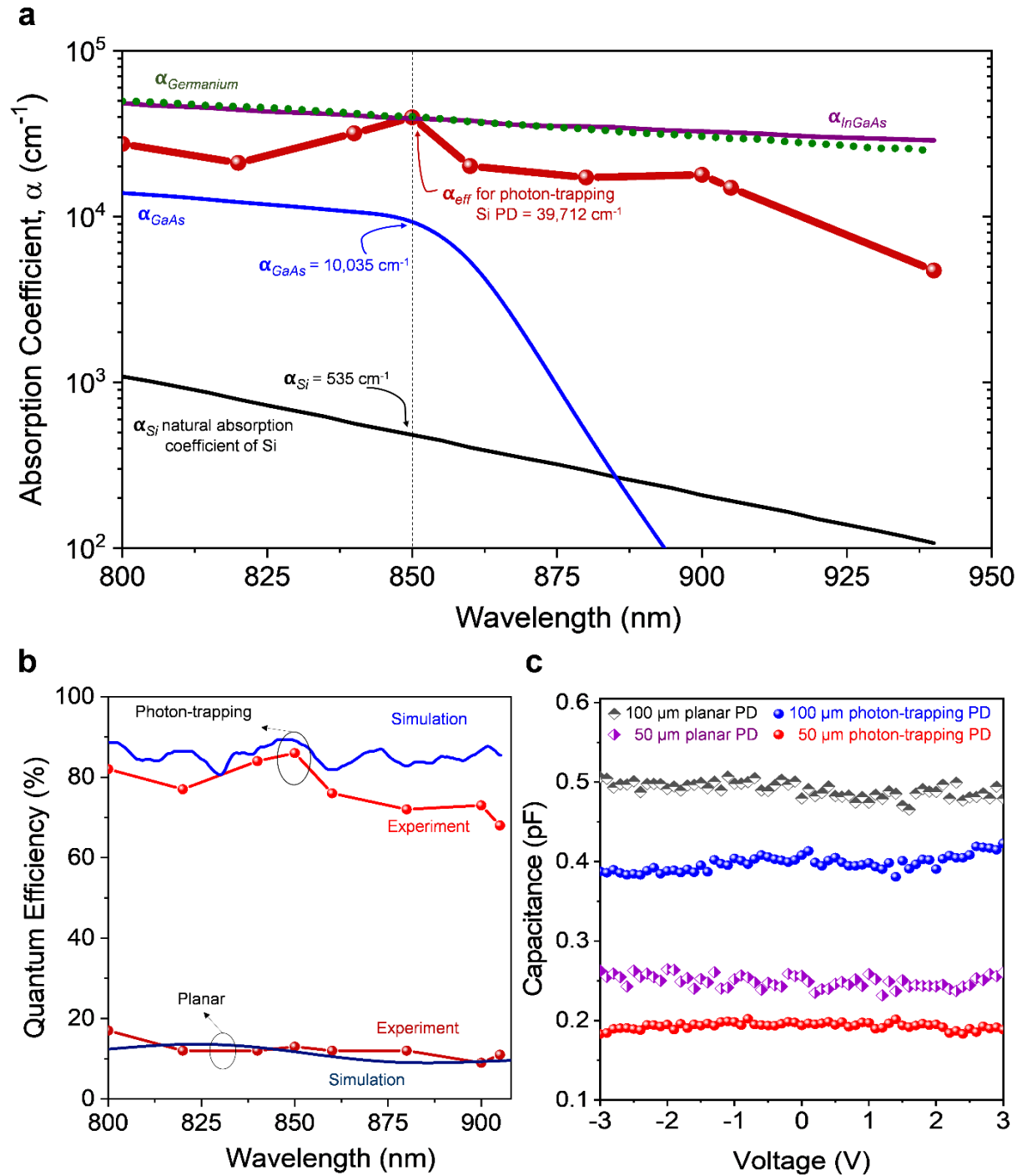

**Fig. 2. Experimental demonstration of absorption enhancement in Si that exceeds the intrinsic absorption limit of GaAs.** (**a**) A comparison of the enhanced absorption coefficient ($\alpha_{eff}$) of the photon-trapping photodetectors and the intrinsic absorption coefficient of Si (bulk) [56], GaAs[56], Ge[55], and In$_{0.52}$Ga$_{0.48}$As[55]. The absorption coefficient of engineered

photodetectors (PD) shows an increase of 20× at 850 nm wavelength compared to bulk Si, exceeds the intrinsic absorption coefficient of GaAs, and approaches the values of the intrinsic absorption coefficient of Ge and InGaAs. (**b**) The measured quantum efficiencies of the devices have an excellent agreement with FDTD simulation in both planar and photon-trapping devices. (**c**) Photon-trapping photodetectors exhibit reduced capacitance compared to their planar counterpart, enhancing the ultrafast photoresponse capability of the device.

Next, we study quantum efficiency (QE), an essential performance metric to quantify the optical sensitivity of the fabricated photodetectors. The measured QE of photon-trapping photodetectors for incident wavelengths ranging from 800 to 905 nm is shown in Fig. 2(b). We have shown a comparison of QEs of planar and photon-trapping photodetectors. External quantum efficiencies over 80% are observed experimentally in the photon-trapping photodetectors for incident wavelengths below 860 nm. Owing to Si's inherent optical material properties, the absorption decreases sharply above 860 nm wavelength, with a minimum value as low as 68% at 905 nm. However, compared to the photodetectors with a planar surface, the absorption efficiency is increased by >550% in photon-trapping photodetectors at 850 nm wavelength. The corresponding responsivities of the photon-trapping devices exhibit over 500 mA/W, as shown in Fig. S7 in the SI. Besides, the minimum enhancement of quantum efficiency in all the fabricated photon-trapping photodetectors compared to the planar devices is at least more than 280% for the wavelength spectrum between 800 and 905 nm. The measured quantum efficiencies also exhibit an excellent agreement with simulated quantum efficiencies in both planar and photon trapping devices as depicted in Fig. 2(b). The detailed simulation study is discussed in the following section and SI. Such a high absorption enhancement directly results from the generation of optical modes propagating laterally due to the integrated photon-trapping surface structures [Movie S2]. Notably, such photon trapping photodetectors exhibit reduced capacitance compared to the planar counterpart due to reduced surface areas caused by the photon-trapping structures, as experimentally characterized and depicted in Fig. 2(c), leading to enhance bandwidth in the device. Such reduction in capacitance can further strengthen the ultrafast response of the devices caused by the thin absorption layer. So far, to the best of our knowledge, such a high photoabsorption enhancement in silicon photosensor is the first experimental demonstration.

**Performance prediction for ultrathin photon-trapping Si photodetectors**

Leaping on the experimental demonstration of extraordinary enhancement in the performance of photon-trapping photodetectors fabricated on 1 µm thin Si, optical simulations are performed by an FDTD method for the most optimized photon-trapping structure with a *p*, *d*, and depth of 1000 nm, 700 nm, and 600 nm, respectively. Again, this should be mentioned here that the best dimension for the fabricated devices was *p* = 1300 nm, *d* = 1000 nm, and depth = 600 nm. The discrepancy could occur due to the limitations of the fabrication technologies [50]. The detail of the optical simulation method is provided in the SI. Such a photon-trapping structure is simulated for absorption in the wavelengths ranging from 800 to 1100 nm as presented in Fig. 3(a). Additionally, a silicon slab with a 1 µm thickness and a planar surface is also simulated as a reference. The red curve represents simulated absorption spectra of the photon-trapping structure for normally incident light. The photon-trapping structure exhibits distinctly higher absorption in comparison with the planar structure. For Si structure with photon-trapping holes and the normally incident light illumination, a maximum photon absorption exceeding 85% is achieved around 850 nm wavelength. The photon-trapping structures facilitate absorption enhancement through guided lateral modes for a broad range of NIR wavelengths. Fig. 3(b) and 3(c) represent the calculated Poynting vector in the photon-trapping silicon slab with 1 µm thickness on the x-y and x-z planes. The figure demonstrates how an ensemble of integrated holes induces a change in the direction of the propagating photons from vertical to lateral orientation in Si films. Laterally oriented Poynting vectors form vortex-like circulation patterns around the sidewalls of the cylindrical holes, resulting in guided light propagation parallel to the photosensor surface for a prolonged time and enabling absorption in Si with high efficiency.

Notably, the guided lateral modes in the Si active layer are also facilitated by the front and the back air/Si and Si/oxide layer interfaces, where the oxide layer of the SOI acts as a back reflector. It should be explicitly mentioned here that the oxide layer significantly facilitates the high absorption of photons in the active layer [50]. A movie demonstrating the side view of energy density distribution mapping of a normally incident light beam that bends almost at a right angle into laterally propagating modes of light along the plane of the Si absorber layer with photon-trapping holes is provided in the SI [Movie S2]. To summarize, the photon-trapping surface structures increase the optical path length, which improves the absorption efficiency within the structure with the aid of enhanced light-matter interactions.

To study and analyze the photon absorption limits of ultra-thin Si films used in modern CMOS processes, we further explored the absorption efficiency of 30 and 100 nm ultra-thin Si films integrated with and without photon-trapping structures, as depicted in Fig. 3(d). Similar to the Si film with 1000 nm thickness, photon-trapping ultra-thin Si film exhibits dramatically higher absorption efficiency than the planar Si film. This also proves that such enhancement in absorption is a direct consequence of enhanced light-matter interaction, irrespective of the film thickness, as visualized by a movie provided in the SI (Movie S3). More than 21% and 8% absorption efficiencies are observed at 850 nm illumination wavelength in Si with 100 and 30 nm absorption thickness, respectively. By contrast, the absorption efficiency is less than 1% for such ultrathin planar Si, as depicted in Fig. 3(d). The estimated $α_{eff}$ for 100 and 30 nm Si layers are 23,572 and 27,794 cm$^{-1}$, respectively, which are significantly higher than the intrinsic absorption limit of GaAs at 850 nm wavelength.

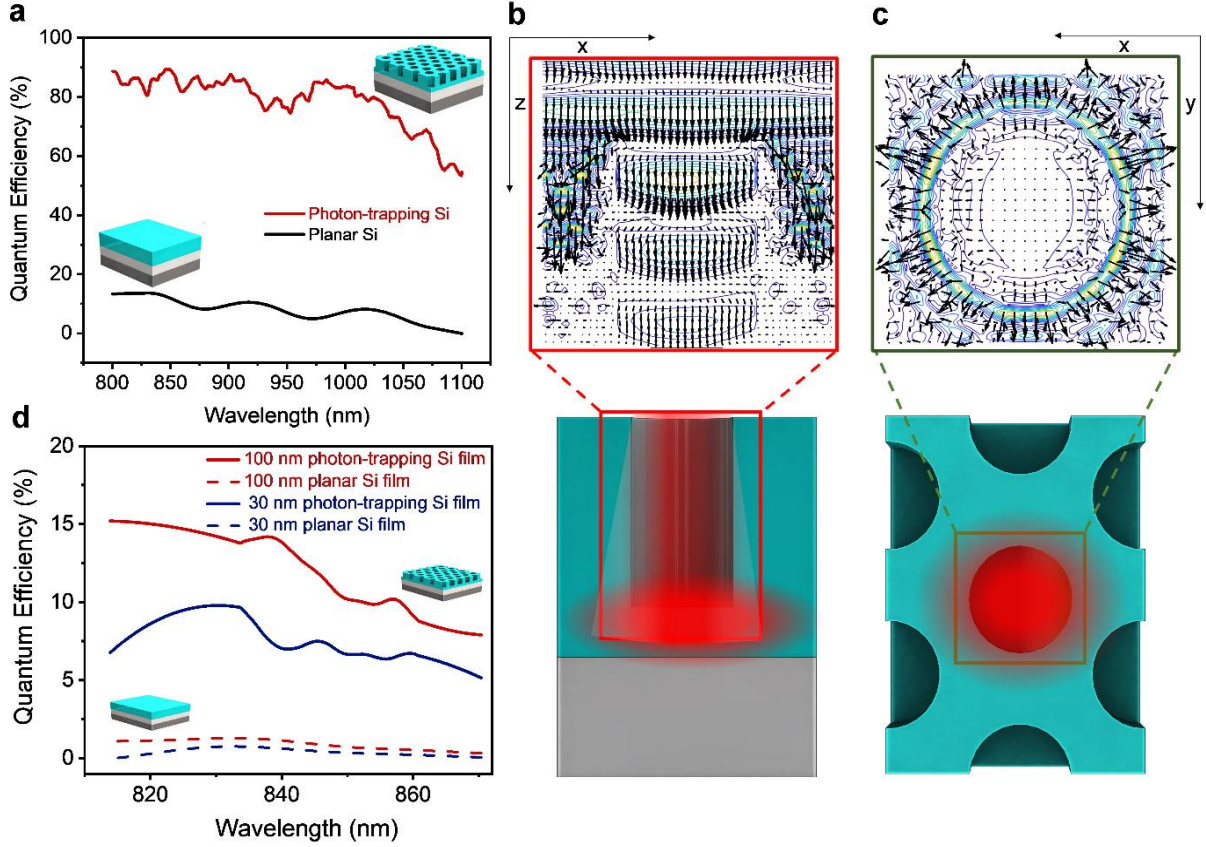

**Fig. 3. Theoretical demonstration of enhanced absorption characteristics in ultra-thin silicon film integrated with photon-trapping structures.** (**a**) A comparison of simulated absorption of photon-trapping (Fig. 1a and S6) and planar structures demonstrates absorption efficiency in photon-trapping silicon around 90% in 1 μm thickness. In contrast, the black curve shows extremely low absorption efficiency in planar silicon without such surface structures. Calculated Poynting vector in holey 1 μm thin silicon on (**b**) x-z (cross-section) and (**c**) x-y (top-view), planes showing that the vectors originated from the hole and moved laterally to the silicon sidewalls, where the photons are absorbed. (**d**) Simulated enhanced optical absorption in ultrathin silicon of 30 and 100 nm thickness with and without photon-trapping structures.

Intriguingly, we also observed that the enhanced absorption coefficient of our fabricated devices effectively exceeds the $4n^2$ limit, as provided in the SI (Fig. S8), where $n$ is the refractive index of silicon at the corresponding wavelengths. Nevertheless, we cannot claim that it exceeds the light trapping geometrical limit of $4n^2$ [15,19], since a collimated laser beam was used for device illumination. In contrast, an isotropic and incoherent light source is commonly used in solar cell characterization. For the case of a collimated beam, the geometrical limit is adjusted by $4n^2/(\sin\theta)^2$, where $\theta$ is the angle between the light source and a plane perpendicular to the surface [27]. We further noticed that the absorption enhancement of our devices could reach a maximum up to $70n^2$ limit at 850 nm of incident wavelength (Fig. S8). Further investigation correlating the illumination angle of the collimated beams and absorption enhancement will help understand if the light trapping geometrical limit can be overcome.

**A physics-based explanation for the performance enhancement**

Based on the above observations, photon-trapping structures effectively supporting lateral modes and efficient coupling of light are essential for the optimum enhancement of absorption efficiency. The eigenmodes of the micro-hole array determine the propagation of photons in the lateral direction. The calculated band structures and allowable available eigenmodes with small holes ($d$=100 nm, $p$= 1000 nm, and thickness, $t_{Si}$ = 1000 nm) and large holes ($d$=700 nm, $p$= 1000 nm, and thickness, $t_{Si}$ = 1000 nm) are shown in Figs. 4(a), 4(b), and S10 for the lateral light propagation in a thin film with an array of holes with the period, $p$, and hole size, $d$. The band structure is calculated using the standard technique that converts Maxwell's equations from *(r, t)* space into *(k,ω)* space by solving the number of wavevectors $k=mp$, where *m* is an integer number. The eigenmodes in the array were calculated at wavelengths near 850 nm, while the ratio of hole diameter and period of the photon-trapping photodetectors is assumed to be *d/p≈0.7* to make it consistent with the fabricated devices. The number of eigenmodes for the hole-array structures increases with an increasing value of *p/λ*, where *p/λ* is larger than unity. Figures 4(a) and 4(b) illustrate the relationship between lateral wavevector *k* and the incident wavelength for a hole size of 100 and 700 nm, respectively, under TE and TM polarizations. The solid curves represent the solutions for ΓX and ΓM directions, whereas the dots represent the areas between those directions, which are also provided in Fig. S10. It is noticeable that the larger holes pronounce curves with a smaller slope, which corresponds to a smaller group velocity. The next question is the coupling into the array. When significant lateral field components are generated in the cylindrical coordinates for a hole with a specific dimension, the solutions of the wave vectors using Bessel functions can be given as, $k_0 = 2\pi/\lambda$, $q_1^2 = k_0^2 - \beta^2$, and $q_2^2 = \varepsilon k_0^2 - \beta^2$, where $k_0$ is the wave vector for a given frequency in the vacuum, $\varepsilon$ is the dielectric constant of silicon, and $\beta$ is the propagation constant. When the solution to the Bessel function $q_2$ is $k_c$, the lateral wave vectors are coupled with eigenmodes. The cross of the solution $k=k_c$ or $k=0$ with the eigenmodes pronounces the modes that propagate laterally and can be absorbed in the material. It should be noted that the Bessel function was solved for a single hole, and it is expected that a similar characteristic can be achieved for an array of holes. For the small holes, we have the solutions only for the finite number of the eigenmodes with $k=0$, which corresponds to the guided modes in photonic crystals. Such structures exhibit sharp spikes in absorption, as shown in Fig. 4(e). However, a continuous solution can be found in the large holey structures for the wavelengths ranging from 800 to 1000 nm, leading to a distinctly higher light absorption than in small holes, as depicted in Fig. 4(e). Hence, the larger holey structures exhibit a good coupling phenomenon due to the relationship between the $k$ vector and the eigenmodes.

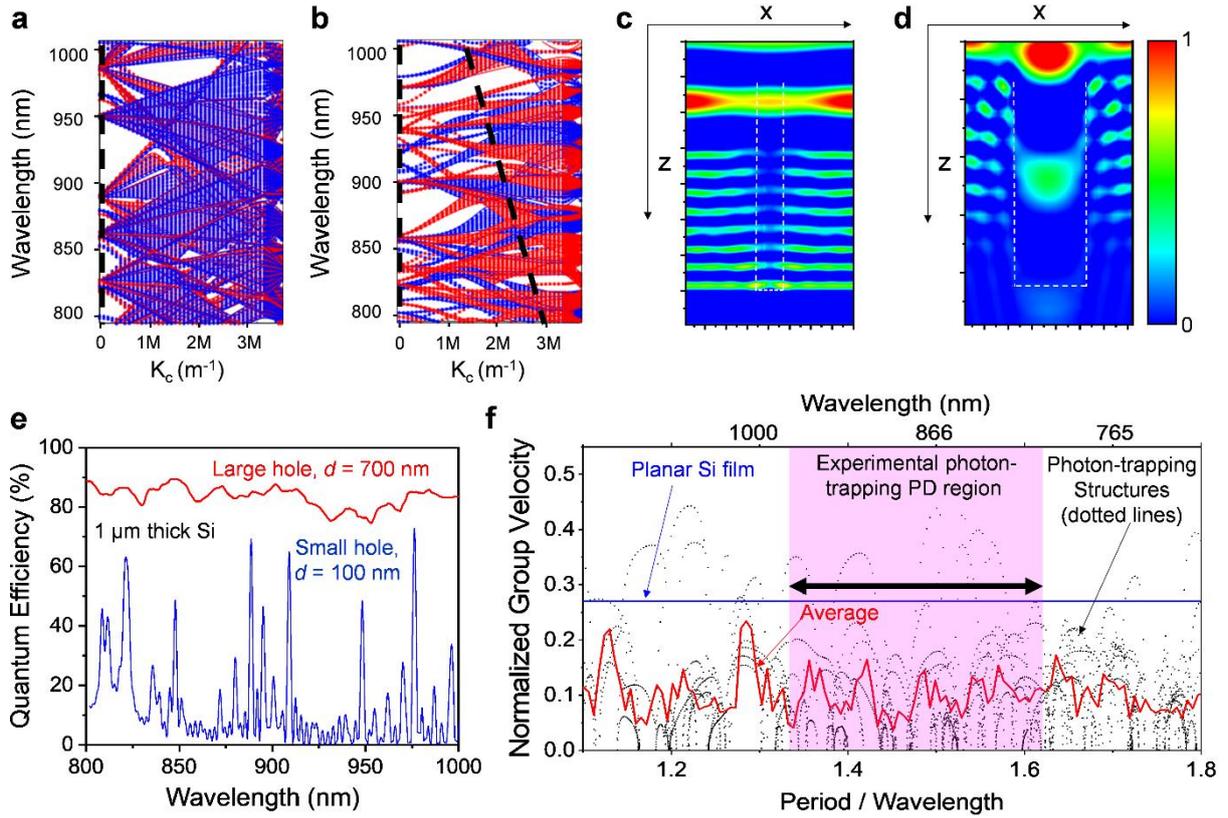

**Fig. 4. Reduced group velocity in photon-trapping silicon (slow light) and enhanced optical coupling to lateral modes contribute to enhanced photon absorption.** Calculated band structure of silicon film with (**a**) small holes ($d$=100 nm, $p$= 1000 nm, and thickness, $t_{Si}$ = 1000 nm) and (**b**) large holes ($d$=700 nm, $p$= 1000 nm, and thickness, $t_{Si}$ = 1000 nm). Red curves represent TE and blue curves represent TM modes. Slanted dash lines are solutions for $k_c$ that couple into the lateral propagation for a vertically illuminating light source. Small hole structures exhibit solutions only for the finite number of the eigenmodes with $k=0$ (vertical dashed line), while large hole structures essentially have both solutions $k=k_c$ and $k=0$ (vertical and slanted dash lines) with the eigenmodes, pronouncing enhanced coupling phenomena and laterally propagated optical modes. (**c, d**) FDTD simulations exhibit optical coupling and the creation of lateral modes. Low coupling and photonic bandgap phenomena are observed for the hole size smaller than the half-wavelength (**c**). Larger holes that are comparable to the wavelengths of the incident photons facilitate a higher number of optical modes and enhanced lateral propagation of light (**d**). (**e**) Calculated optical absorption in silicon with a small hole ($d$=100 nm, $p$= 1000 nm, and thickness, $t_{Si}$= 1000 nm) compared with the absorption of the large hole ($d$=700 nm, $p$= 1000 nm, and thickness = 1000 nm). (**f**) For frequencies (period of holes/light wavelength) between ~1.3 - 1.6, the normalized light group velocity (red curve) for 850nm wavelength is significantly lower in photon-trapping silicon compared to that of the bulk Si (blue line). The red curve represents an averaged group velocity for Si photon-trapping structures which exhibits a distinctly lower value in our fabricated devices.

In the next step, the influence of the size of the holes on the formation of lateral optical modes and the corresponding field distribution is studied. Coupling phenomena are only observed in the photon-trapping structures, as presented in Fig. S11. Low coupling phenomena are observed for the devices with hole sizes smaller than half the wavelength as shown in Fig. 4(c), where photons cannot efficiently couple within the absorber layer. However, for the hole size comparable with the incident wavelength, the light can couple into the holes and leak out through the sides of the hole, as illustrated in Fig. 4(d). The incident photons also reflect from the surface of photodetectors when the hole diameter is smaller than the incident wavelength, which is not the case for our most efficient fabricated devices. The hole diameters of our most optimized fabricated devices are comparable to the incident wavelength, similar to the one shown in Fig. 4(d). In photodetectors with larger holes, photons accumulate in the x-z plane around the hole after coupling into laterally propagating modes and eventually getting completely absorbed there. Furthermore, the vertically illuminated light refracts at *atan(n)* angle from the boundary conditions, increasing light absorption in the active layer. Finally, the oxide layer of the SOI wafer underneath the sensors further contributes to enhanced photon absorption by reflecting the photons in the direction of the device surface. The influence of coupling for smaller and bigger holes at different incident angles is also studied on the photon absorption as provided in the SI (Fig. S4). photodetectors with larger holes (d/p~0.77) exhibit noticeably higher light absorption irrespective of incident angles than devices with smaller holes. However, when illuminated with photons of longer wavelengths, a relatively higher photon absorption is obtained for the incident angle of 30° compared with 0° in both small and larger holes.

Slow light with reduced group velocity increases absorption efficiency due to the augmented light-matter interactions. The group velocity was calculated from the band diagram as $u_g=d\omega/dk$ under TE polarization modes, as presented in Fig. 4(f). The group velocity in bulk

Si was calculated as *c/n*, where *c* is the light velocity and *n* is the refractive index of Si at 850 nm. The normalized frequency of our experimental structures is between 1.3 to 1.6. Herein, the group velocity in most modes for the normalized frequency between 1.3- 1.6 is significantly lower than the group velocity of light in bulk Si. The average group velocity for the modes was also calculated, as shown by the red line, indicating a conspicuously lower average group velocity of the photon trapping structures compared to that of Si without such surface structures. Light-trapping surface structures have been demonstrated to be capable of enabling enhanced optical density of states (DOS) with light enhancement beyond the ray optic limit [57,58]. Herein, the DOS was also calculated as an integral over the wavevector for a given frequency, $\rho(\omega) = \int \left\{\frac{d^d k}{(2\pi)^d}\right\} \delta(E - E(k))$, *d* and *E* are differential dimension and energy, respectively. For the 2D photonic crystals, DOS could be approximated as $\rho_{2D}(\omega) = \frac{4\epsilon\omega}{\pi c^2}$ [27]. The DOS of the photon-trapping photodetectors is found to be higher than the photodetectors with planar surfaces, as provided in Fig. S9(a). The photodetectors with micro-hole periods shorter than the incident wavelengths exhibit noticeably low DOS compared to the periods comparable to and slightly longer than the wavelengths. Our designed and fabricated photodetectors closely match with the shaded area on the right half of Fig S9(a), exhibiting high DOS for frequencies higher than 1.0 (periods are comparable to or slightly longer than the incident wavelengths). Nevertheless, the region with high values of *p/λ* needs more Fourier components, and the maximum peaks of the DOS are not as pronounced as expected, while there is a constant increase of the optical mode density with *p/λ*. We conclude that observed absorption enhancement in Si is a combined effect of slow light with reduced group velocity, the enhanced photonic DOS, lateral propagation of a large number of optical modes, and efficient coupling of incident photons to the photon-trapping structures integrated on the Si surface. The cumulative impacts of the aforementioned processes help Si enhance light absorption by more than twenty-fold and exceed the intrinsic absorption limit of GaAs.

We used a one-micron thin silicon film with photon-trapping structures to demonstrate ultrafast time response characteristics of ~31 picosecond (ps) full width at half maximum (FWHM) and 16 ps deconvolved time response at 850 nm, as shown in Figure S9(b). Such ultra-fast time response coupled with broadband very high absorption efficiency exhibited by ultrathin Si helps overcome the bandwidth-absorption trade-off faced by the ultra-fast photodetector community. This work, thus, is very relevant to the design and fabrication of extremely fast and highly sensitive integrated photodetectors using modern CMOS foundry processes that currently use ultrathin Si of similar thickness.

**Conclusions**

This paper presents an experimental demonstration of photoabsorption enhancement by more than 20× in silicon that effectively exceeds the intrinsic absorption limit of GaAs for a broad wavelength spectrum between 800 to 905 nm. In our demonstration, we used photodetectors designed with a 1 µm thin silicon film integrated with an array of periodic photon-trapping structures that allow the generation of slow light with reduced group velocity, enhanced photon density of states (DOS), and, consequently, very high density of laterally propagating slow and stationary optical modes. This results in a more prolonged light-matter interaction time within the thin absorption region of the device. Intriguingly, simulated performances for ultrathin silicon photodetectors with even thinner absorption regions, such as 30 and 100 nm thin films, exhibit similarly enhanced photosensitivity. Additionally, Si photon-trapping structures help reduce photodetectors' capacitance compared to the planar counterpart, enabling faster response. The fabrication process is CMOS compatible and can contribute to integrated photodetectors with ultrafast responses for data communication systems, emerging biomedical imaging applications, biosensing, and autonomous vehicles.

**Acknowledgments**
We thank Eli Yablonovitch for valuable discussions and H. Cansizoglu and Y. Gao for their help with the device fabrication. **Funding:** This work was supported in part by the US Army's Night Vision and Electronic Sensors Directorate under Grant # W909MY-12-D-0008, NSF ECCS grant # 1428392, and by the S. P. Wang and S. Y. Wang Partnership, Los Altos, CA. Cesar Bartolo-Perez acknowledges the National Council of Science and Technology (CONACYT) and UC-MEXUS for the Doctoral fellowship. Part of this study was carried out at the UC Davis Center for Nano and Micro Manufacturing (CNM2).